{{}}\documentclass[twoside,letterpaper]{pazh_mod}
\usepackage{graphicx}
\usepackage{natbib}

\begin{document}

{{}}\journalinfo{2007}{00}{0}{001}[010] 

\title{Hydrogen-like nitrogen radio line from
       hot interstellar and warm-hot intergalactic gas}
\date{}


{{}}\author{%
{{}}  R.A.\ Sunyaev\address{1,2}, 
{{}}  D.\ Docenko\address{1}\email{dima@mpa-garching.mpg.de}
{{}}  \addresstext{1}{Max-Planck-Institut f\"ur Astrophysik,
{{}}           Karl-Schwarzschild-Str. 1,
{{}}           Postfach 1317,
{{}}           85741 Garching,
{{}}           Germany}
{{}}  \addresstext{2}{Space Research Institute, 
{{}}                  Russian Academy of Sciences, 
{{}}                  Profsoyuznaya ul. 84/32, Moscow 117810, Russia}
{{}}}

{{}}\shorttitle{14-N VII LINE IN SPECTRA OF ISM AND WHIM}
{{}}\shortauthor{SUNYAEV \& DOCENKO} 



\begin{abstract}
Hyperfine structure lines of highly-charged ions may open
a new window in observations of hot rarefied astrophysical plasmas. 
In this paper we discuss spectral lines of isotopes and ions 
abundant at temperatures $10^5-10^7$~K, characteristic for 
warm-hot intergalactic medium,
hot interstellar medium,
starburst galaxies, their superwinds and 
young supernova remnants. 
Observations of these lines will allow to study 
bulk and turbulent motions of the observed target 
and will broaden the information about the gas ionization state,
chemical and isotopic composition.

The most prospective is the line of the major nitrogen isotope
having wavelength $\lambda=5.65$~mm~\citep{SCh84}.
Wavelength of this line is well-suited for observation of 
objects at $z\approx0.15-0.6$ when it is redshifted to $6.5-9$~mm 
spectral band widely-used in ground-based radio observations,
and, for example, for $z\ge 1.3$, when the line can be observed in
1.3~cm band and at lower frequencies.
Modern and future radio telescopes and interferometers 
are able to observe the absorption by $^{14}$N~VII 
in the warm-hot intergalactic medium 
at redshifts above $z\approx0.15$ 
in spectra of brightest mm-band sources.
Sub-millimeter emission lines of several most abundant isotopes
having hyperfine splitting 
might also be detected in spectra of young supernova remnants.



\end{abstract}


\section{Introduction}
\label{SecIntro}

In the temperature range $10^5-10^7$~K 
the most abundant ion having hyperfine splitting of the ground state
is the H-like ion of the major nitrogen isotope $^{14}$N~VII
with line
in radio range at around 5.64~mm~\citep{SCh84}. 
This wavelength of transition between its 
hyperfine structure (HFS) components has been estimated with high 
precision to be $\lambda=5.6519(11)$~mm~\citep{Shabaev95}. 
Wavelength of this line is well suited for observation of 
objects at $z\approx0.15-0.6$ when it is redshifted 
to $6.5-9$~mm spectral band widely-used in ground-based radio observations,
and, for example, for $z\ge 1.3$, when the line can be observed in
1.3~cm band and at lower frequencies. Unfortunately, for our Galaxy
and its vicinity with $z<0.1$ it is difficult to observe this line due to
atmospheric absorption.
Possibly, this is the main reason why there have been no attempts to
observe it from the ground after this line was reported by~\citet{SCh84}
as a promising candidate for detection of hot rarefied plasmas.
Till now astrophysical plasma at such temperatures has been studied 
only by rocket-based instruments and space ultraviolet and 
soft X-ray missions such as 
\textit{ROSAT}, 
\textit{FUSE}, 
\textit{Suzaku}, 
\textit{Chandra} 
and \textit{XMM-Newton}. 

One of important and interesting predictions of large-scale
structure simulations using hydrodynamical approach is the existence of
rarefied intergalactic gas heated to temperatures $T\approx10^5-10^7$~K.
This warm-hot intergalactic medium (WHIM, 
e.g., \cite{CenWhere1} and \cite{CroftWHIM,Dave01}) contains dominant fraction
of barions in the present Universe (according to the cosmic census 
of barions made by \cite{FukugitaBudget}), but
is practically unobserved till now. 
Computations by~\cite{Gnedin,CenWhere1} show that heavy element
abundances in it rise sharply in regions of higher temperature and
density reaching values close to the solar ones.

There were many attempts to observe the WHIM at $z>0$, but till now only
far-ultraviolet absorption lines of lithium-like oxygen
($\lambda\lambda=1032,1038$~\AA) have been detected by 
\textit{Hubble Space Telescope} 
for $z>0.12$ and by 
\textit{FUSE} spacecraft for $z<0.15$ (e.g.,~\cite{DanforthOVI}).
Despite many attempts using \textit{Chandra} and
\textit{Suzaku}~\citep{SuzakuA2218}
orbital telescopes, O~VII and O~VIII soft X-ray lines for $z>0$ are evading
detection till now (for example, reported case of detection in~\cite{Nicastro}
was recently criticized by~\cite{NicastroAnti1} and~\cite{NicastroAnti2}).
Realization of planned proposals considering the micro-calorimeter
X-ray sky surveys (e.g.,~\cite{SMEX}) 
will certainly highly increase the probability to
detect soft X-ray emission and absorption lines from the WHIM.

It is obvious that existence of ground-based methods of detection 
of this gas will bring a lot of additional strength in such efforts.
We are proposing radio observations of hyperfine structure
lines of highly-charged ions in sub-millimeter to centimeter bands
as a way to get additional information about 
the velocity field, mass, temperature and chemical 
abundance distribution of the warm-hot intergalactic medium.

It is important to mention that spectral resolution of radio detectors
is better than even that of micro-calorimeters.
Therefore these lines might permit to look for turbulence and 
bulk motions in the objects of interest~\citep{KKHP-SK}.
In addition, radio methods do not only probe
the absorption lines in spectra of brightest radio sources 
in millimeter spectral band, but also the corresponding 
emission of the gas.

None of the lines we are discussing was detected so far neither
in astrophysical objects, nor in physical laboratory conditions.
Simplest estimates using well-known relativistic corrections and
approximations~\citep{Sobelman1}
give precision of the HFS line wavelength on the order
of few percents. The list of hyperfine structure lines of interest 
for study of hot plasmas, published by~\cite{SCh84}, attracted
attention of atomic physicists.
\cite{ZhangSampson97,ZhangSampson00} made computations
of electronic excitation of HFS levels in plasmas using 
relativistic distorted-wave method, accounting for resonance effects. 
\cite{Shabaev95,Shabaev97} calculated more
precise transition wavelengths using a combination of configuration
interaction Hartree-Fock method and the $\frac1Z$ perturbation theory.

Observations of supernova remnants brightest in soft X-ray band 
(according to \textit{ROSAT} All-Sky Survey~\citep{RASS-BSC} 
and recent \textit{XMM-Newton} studies)
indicate the most prospective objects to search for HFS
line emission. Knowledge of the studied object radial velocity 
from optical observations will allow to measure wavelengths of 
the lines with precision of at least
10~km/s (corresponding to wavelength uncertainty of
30~ppm) that is better than the theoretical estimate precision 
(200~ppm or more). 
Knowledge of exact wavelengths will also be very important for study 
and identification of absorption lines from WHIM filaments.
Additionally, one must not forget that such theoretical computations
should always be confirmed by experimental measurements.

Among such brightest targets we made estimates of brightness of 
particular HFS lines arising in 
Cygnus Loop, North Polar Spur, Vela XYZ, N157B and Cas~A 
supernova remnants (SNR). Our estimates show that in all
these objects (except for too hot Cas~A) the line of $^{14}$N~VII is
the brightest. Unfortunately, it is subject to atmospheric 
obscuration.

At Chajnantor plain height ($h\approx5100$~m, where \textit{APEX}
telescope is operating and 
\textit{Atacama Cosmology Telescope} and 
\textit{ALMA} interferometer are being built)
the air specific attenuation is already significantly less 
than on the sea level (0.2 vs.\ 1.1 dB/km at 53~GHz,~\cite{Liebe60GHz})
and atmospheric transmission of $^{14}$N~VII line may be as high as 
30--50\%~\citep{SchwabHogg}.
This opens a possibility of direct detection of nitrogen radio line 
from bright supernova remnants of the southern skies and the Cygnus Loop.
High-altitude radio telescopes, naturally, are able to observe this 
line also at much lower redshifts than the instruments at the sea level.

There are many well-studied star-forming galaxies at redshifts $z=0.2-0.5$
which should have a lot of relatively young supernova remnants 
with gas in interesting range of temperatures. Such SNRs and galaxies as 
a whole would be extremely interesting objects to study by means of 
hyperfine structure lines using existing (such as
\textit{Green Bank Telescope} 
and \textit{VLA}) 
and future (such as 
\textit{Square Kilometer Array} 
and \textit{ALMA}) 
radio telescopes and interferometers.
Hyperfine lines might be bright also in strong outflows of the hot gas 
from star-forming galaxies (e.g.,~\cite{GalOutflows}). It is commonly believed 
that such galactic winds is one of the ways
of the intergalactic medium heavy element enrichment.

\bigskip
Hyperfine line observations in principle permit to separate 
contributions of heliospheric charge-exchange emission and of
the Local Hot Bubble, that is now constituting a hardly solvable 
problem for soft X-ray measurements (e.g.,~\cite{MBM12Chandra}).
This becomes possible due to HFS line emissivity dependence 
on the plasma density (see below), and the fact that
the solar wind is much denser than the gas of the Local Bubble.

The Galaxy is essentially transparent in millimeter band. 
Therefore the HFS lines allow studying objects that are
strongly obscured in visible and soft X-ray bands. 
Line emission will mostly probe young supernova remnants, but the 
total absorption column in HFS lines is a measure of hot 
($T\approx10^6$~K) interstellar gas
in the Galaxy, thus being complimentary to H~I 21~cm line probing
neutral interstellar medium (ISM).

In case of observations of young supernova remnants where the bulk 
of emission is coming from the enriched material, it becomes possible
to measure the isotopic composition by comparing the HFS line
intensity (sensitive to just one isotope) with X-ray
lines of the same element (produced by all isotopes).

\bigskip
The structure of the paper is the following. In the first part 
we are presenting wavelengths and other data on 
the most abundant isotopes in the temperature range $10^5-10^7$~K having hyperfine
structure lines in sub-millimeter, millimeter and centimeter bands.
We are discussing sub-millimeter lines in connection with \textit{ALMA}
array 
being able to observe in this spectral band.
We are also computing the HFS level population as a function 
of electron density and radiation temperature for different transitions. 
In the second part
we are presenting the results of the emission line differential brightness 
temperature computations from some of the brightest objects 
in our Galaxy and its surroundings. In the last Section we are 
presenting our estimates 
of HFS absorption line optical depth arising in WHIM and hot ISM.

\section{Hyperfine structure transitions}
\label{SecData}

\begin{figure*}[htb]
  \begin{center}
{{}}  \centerline{ 
    \rotatebox{270}{
{{}}       \includegraphics[height=0.49\linewidth]{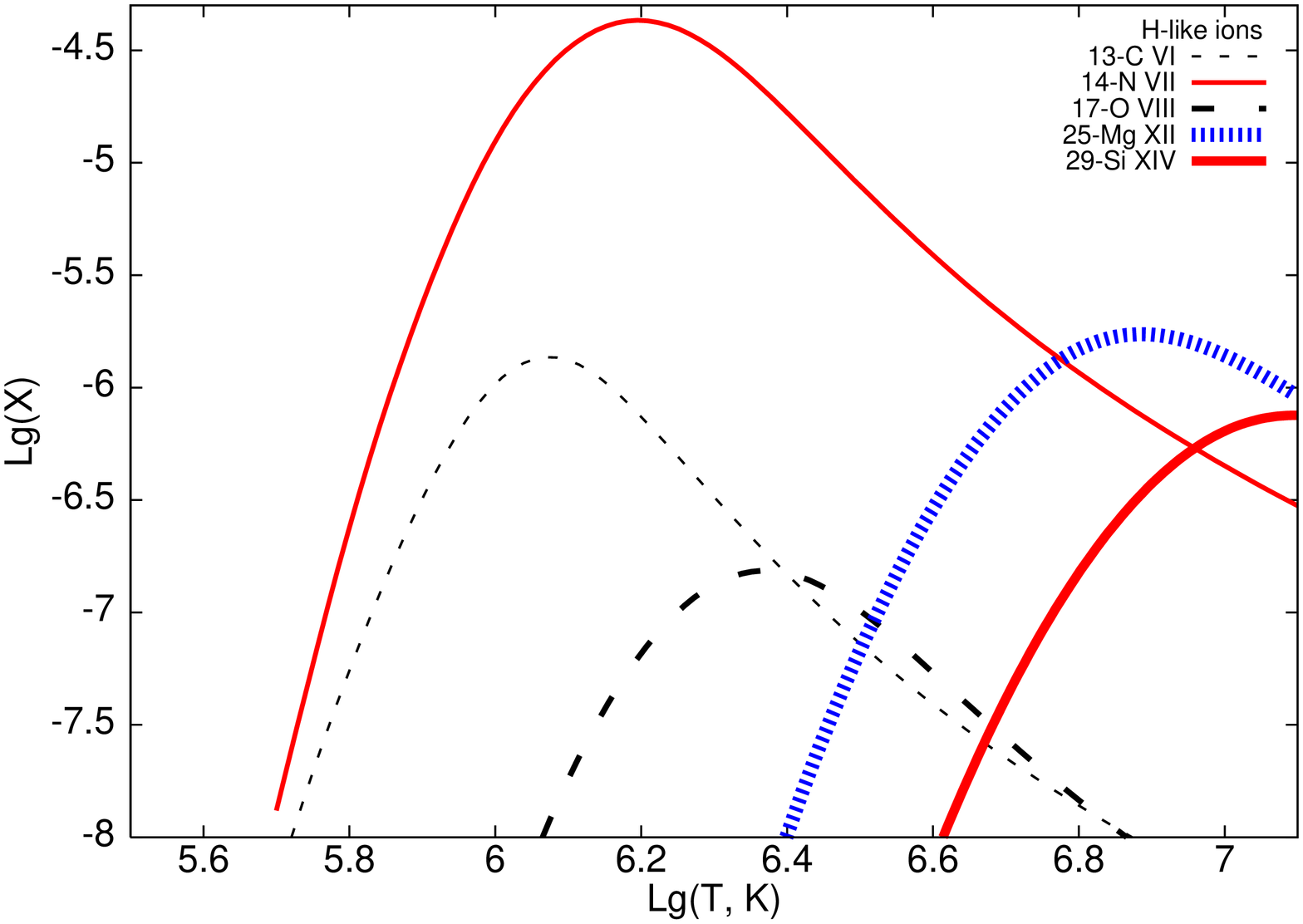}
                   }
    \rotatebox{270}{
{{}}       \includegraphics[height=0.49\linewidth]{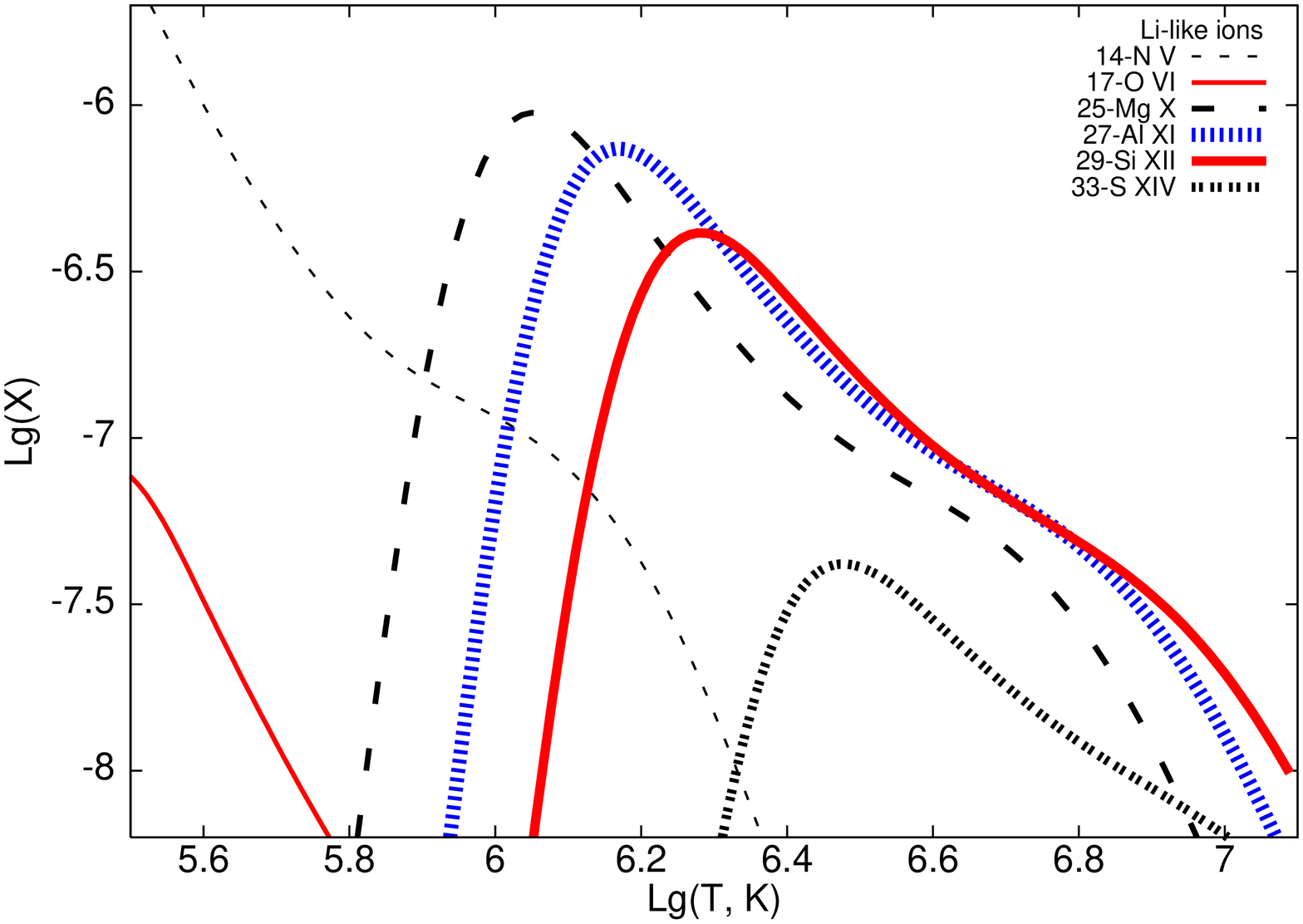}
                   }
{{}}             }
  \caption{Abundance $X=X_{ion}(T)\cdot X_{iso}$ of discussed ions of isotopes
   at WHIM, SNR and hot ISM temperatures. Collisional ionization
   equilibrium ionic fractions $X_{ion}(T)$, solar elemental abundances 
   and Earth isotopic mole fractions 
   are assumed. Left and right panels present data on 
   H-like ions and Li-like ions respectively.
   Note the difference in the vertical scale.}
   \label{FigHFSabund}
  \end{center}
\end{figure*}

Isotopic abundances $X_{iso}$ (a product of 
solar elemental abundances from~\cite{GS98} and 
Earth-measured isotopic mole fractions by~\cite{NISTiso})
clearly show (see Table~\ref{TabHFSLinesL}) that 
among isotopes having non-zero nuclear spin the $^{14}$N 
is the most abundant after hydrogen. Its hydrogen-like
and lithium-like ions have a hyperfine splitting of the ground 
state and produce hyperfine structure (HFS) line in spectra of
objects with temperatures appropriate for existence of such ions.
We include these ions in our analysis, as well as several less 
abundant H-like and Li-like ions of $^{13}$C, $^{17}$O, $^{25}$Mg,
$^{27}$Al, $^{29}$Si and $^{33}$S (see Figure~\ref{FigHFSabund}).

{{}}

{{}}\begin{table*}[hbt]
\caption{Parameters of transitions between the ground state 
hyperfine sublevels: wavelengths $\lambda$, 
transition rates $A_{ul}$ and absorption cross-sections $\sigma$ for 
velocity dispersion of 30~km/s. 
$X_{iso}$ denotes isotopic fraction relative to hydrogen.
Table also includes nuclear spin $I$ and nuclear magnetic moment
$\mu$ expressed in nuclear magnetons (nm).}
  \label{TabHFSLinesL} 
\begin{center}
\begin{tabular}{|l|c c l|l l l|}
\hline
Isotope, ion & $X_{iso}$         & $I$  & $\mu$, nm  
 & $\lambda$, mm  & $A_{ul}$, s$^{-1}$ &   $\sigma$, cm$^2$\\

\hline
$^{13}$C VI  & $3.5\cdot10^{-6}$ & 1/2 & +0.7024118(14) &
 3.8740(8)  & $4.639(3)\cdot10^{-10}$ &   $6.05\cdot10^{-19}$\\
\hline
$^{14}$N V   & $8.3\cdot10^{-5}$ & 1   & +0.40376100(6)&
 70.72(4)   & $1.018(2)\cdot10^{-13}$ &   $5.39\cdot10^{-19}$\\
\hline
$^{14}$N VII & $8.3\cdot10^{-5}$ & 1   & +0.40376100(6)&
 5.6519(11) & $1.9920(12)\cdot10^{-10}$ & $5.39\cdot10^{-19}$\\
\hline
$^{17}$O VI  & $3.2\cdot10^{-7}$ & 5/2 & -1.89379(9)&
 11.813(7)  & $3.818(7)\cdot10^{-11}$ &   $3.37\cdot10^{-19}$\\
\hline
$^{17}$O VIII& $3.2\cdot10^{-7}$ & 5/2 & -1.89379(9)&
 1.0085(2)  & $6.136(4)\cdot10^{-8}$  &   $3.37\cdot10^{-19}$\\
\hline
$^{25}$Mg X  & $3.8\cdot10^{-6}$ & 5/2 & -0.85545(8)&
 6.680(4)   & $2.111(4)\cdot10^{-10}$ &   $3.37\cdot10^{-19}$\\
\hline
$^{25}$Mg XII& $3.8\cdot10^{-6}$ & 5/2 & -0.85545(8)&
 0.65809(13)& $2.2083(13)\cdot10^{-7}$ &   $3.37\cdot10^{-19}$\\
\hline
$^{27}$Al XI & $2.9\cdot10^{-6}$ & 5/2 & +3.6415069(7)&
 1.2060(7)  & $2.563(5)\cdot10^{-8}$  &   $4.71\cdot10^{-19}$\\
\hline
$^{29}$Si XII& $1.7\cdot10^{-6}$ & 1/2 & -0.55529(3)&
 3.725(2)   & $1.566(3)\cdot10^{-9}$  &   $2.01\cdot10^{-19}$\\
\hline
$^{29}$Si XIV& $1.7\cdot10^{-6}$ & 1/2 & -0.55529(3)&
 0.38165(7) & $1.4557(8)\cdot10^{-6}$  &   $2.01\cdot10^{-19}$\\
\hline
$^{33}$S  XIV& $1.6\cdot10^{-7}$ & 3/2 & +0.6438212(14)&
 3.123(2)   & $1.328(3)\cdot10^{-9}$  &   $5.05\cdot10^{-19}$\\
\hline

\end{tabular}
\end{center}
\end{table*}

Non-relativistic formula of hyperfine splitting energy~\citep{Sobelman1}
of one-electron ion in $ns$~$^2S_{1/2}$ state is (in Rydbergs) 
$$
  \Delta E_{HFS} = \frac83 \frac{\mu}{I} \cdot \frac{\alpha^2Z^3}{n^3}
                   \frac{m_e}{m_p}\cdot(I+1/2)\; Ry,
$$
where $\mu$ is the nuclear magnetic moment, $I$ is the nuclear spin,
$\alpha$ is the fine structure splitting constant, $Z$ is the nuclear
charge, $m_e$ and $m_p$ are electron and proton mass and 
$n$ is the principal quantum number of the level.
Sharp (cubic) dependence on the nuclear charge $Z$ is seen. 
Radiation transition probability depends on $Z$ even more sharply 
(proportional to $\Delta E_{HFS}^3$, therefore to $Z^9$, see below), that
makes HFS transitions in high-$Z$ ions competitive to hydrogen 21~cm
transition even despite their much smaller isotopic abundances.

More precise wavelengths of hyperfine structure transitions 
are taken from \cite{Shabaev95,Shabaev97}, except for H-like ions
$^{25}$Mg~XII and $^{29}$Si~XIV where formulas 
given in \cite{Shabaev94} were utilized directly.
Transition rates are then computed using well-known formula
\citep{Sobelman80} that in case of H-like and Li-like ions takes the form
\begin{equation}
  \label{EqHFSrate}
  \begin{array}{l}
       A(F \to F-1) = \frac{I}{I+1} A(F-1 \to F) = \\[5pt]
      \qquad \qquad = 1.0789\cdot10^{-7} \frac{I}{2I+1} 
             \cdot \left(\frac{\lambda}{\mbox{\footnotesize 1 mm}}\right)^{-3}
      {\rm s}^{-1} ,
  \end{array}
\end{equation}
where $F$ is the largest of the two sublevel total atomic angular momenta
and $\lambda_{\rm mm}$ is the transition wavelength
expressed in mm. Note that in case of negative nuclear magnetic 
moment $\mu$ the sublevel with total angular momentum $F-1$ has energy 
larger than one of $F$.

Absorption cross-section $\sigma$ is \citep{Sobolev}:
\begin{equation}
  \label{EqSigRyb}
  \sigma=\frac{\lambda^2}{8\pi}\frac{g_u}{g_l}\phi(\nu)A_{ul},
\end{equation}
where 
$g_u$ and $g_l$ are hyperfine sublevel weights,
$\phi(\nu)$ is a spectral line profile, normalized such that
$\int_{-\infty}^{\infty}\phi(\nu)d\nu = 1$ and 
$A_{ul}$ is the spontaneous transition rate from the upper hyperfine 
sublevel $u$ to the lower $l$.

Near the center of a spectral line
$\phi(\nu)\approx(\sqrt{\pi}\Delta\nu_D)^{-1}$, 
where the Doppler line width 
$\Delta\nu_D$ is arising due to thermal and turbulent motions of 
particles in plasma 
(thermal motion velocity $v_{th}$ of nitrogen ions at temperature
$10^6$~K is about 34~km/s). Inserting this expression
into equation~(\ref{EqSigRyb}), we obtain:
\begin{equation}
  \label{EqSigma}
  \sigma=\frac{\lambda^3}{8\pi^{3/2}\; v}\frac{g_u}{g_l}A_{ul}.
\end{equation}
It is seen that the absorption cross-section is not dependent directly
on the nuclear charge, as the radiative transition rate dependence on
the wavelength is canceled by the $\lambda^3$ factor.

In Table~\ref{TabHFSLinesL} we present parameters of HFS transitions,
such as wavelengths, transition rates and absorption cross-sections.
Abundances of ions (taken from \cite{Mazzotta})
of isotopes of interest as functions of
temperature are shown on Figure~\ref{FigHFSabund}.

HFS sublevel 
electron collisional excitation collision strength values for
$^{13}$C~VI and $^{14}$N~VII are given by~\cite{ZhangSampson97}. 
For other H-like ions they were obtained by scaling along 
isoelectronic sequence (Beigman, private communication).
Respective electron-impact fine-structure excitation cross-sections for 
lithium-like ions were taken from~\cite{FisherLi}.
Already ~\cite{SCh84} noticed that this is the main channel 
of the ground-state HFS excitation in Li-like ions.

\begin{figure*}[htb]
  \begin{center}

{{}}  \centerline{ 
{{}}       \includegraphics[width=0.49\linewidth]{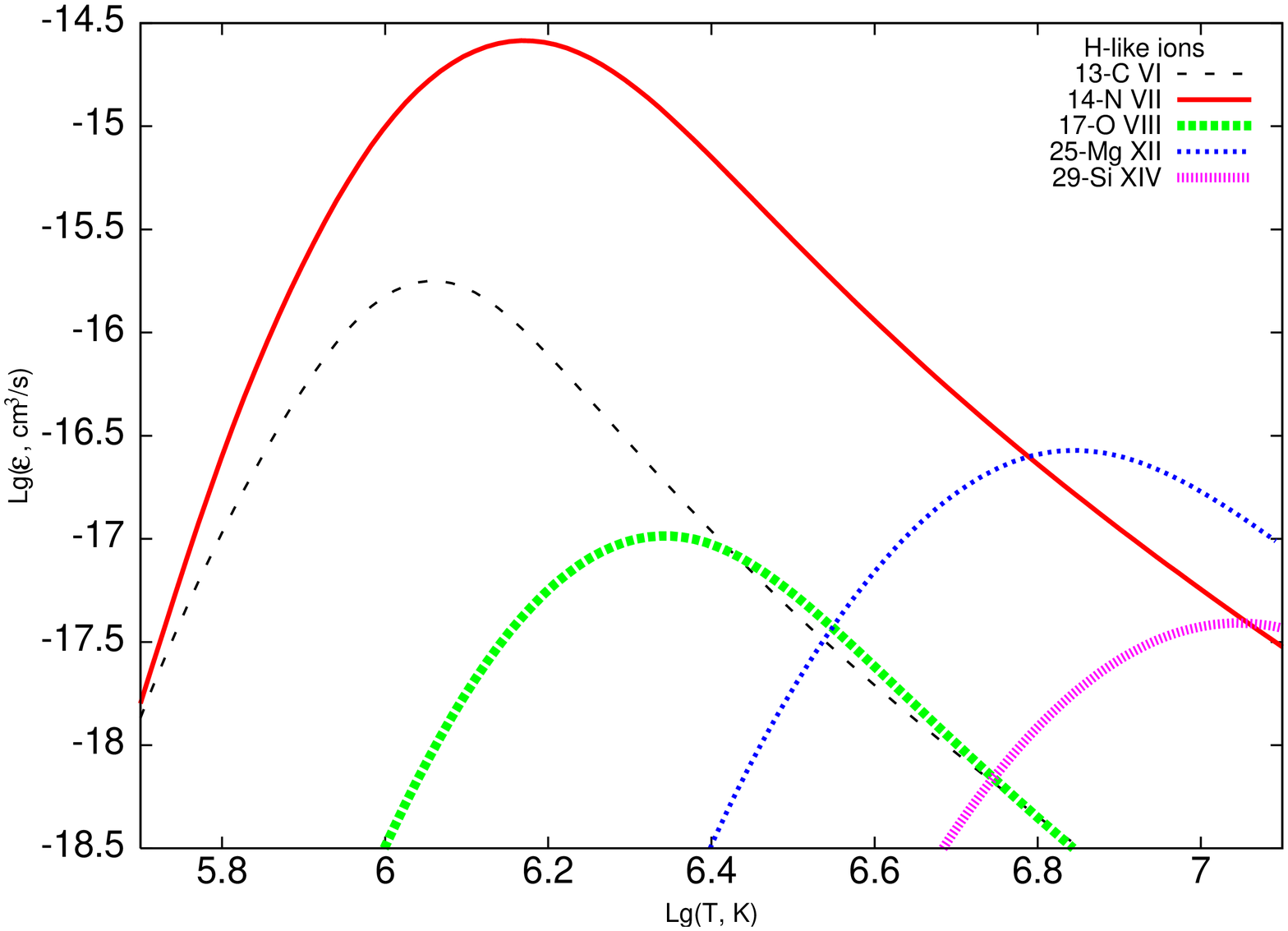}
{{}}       \includegraphics[width=0.49\linewidth]{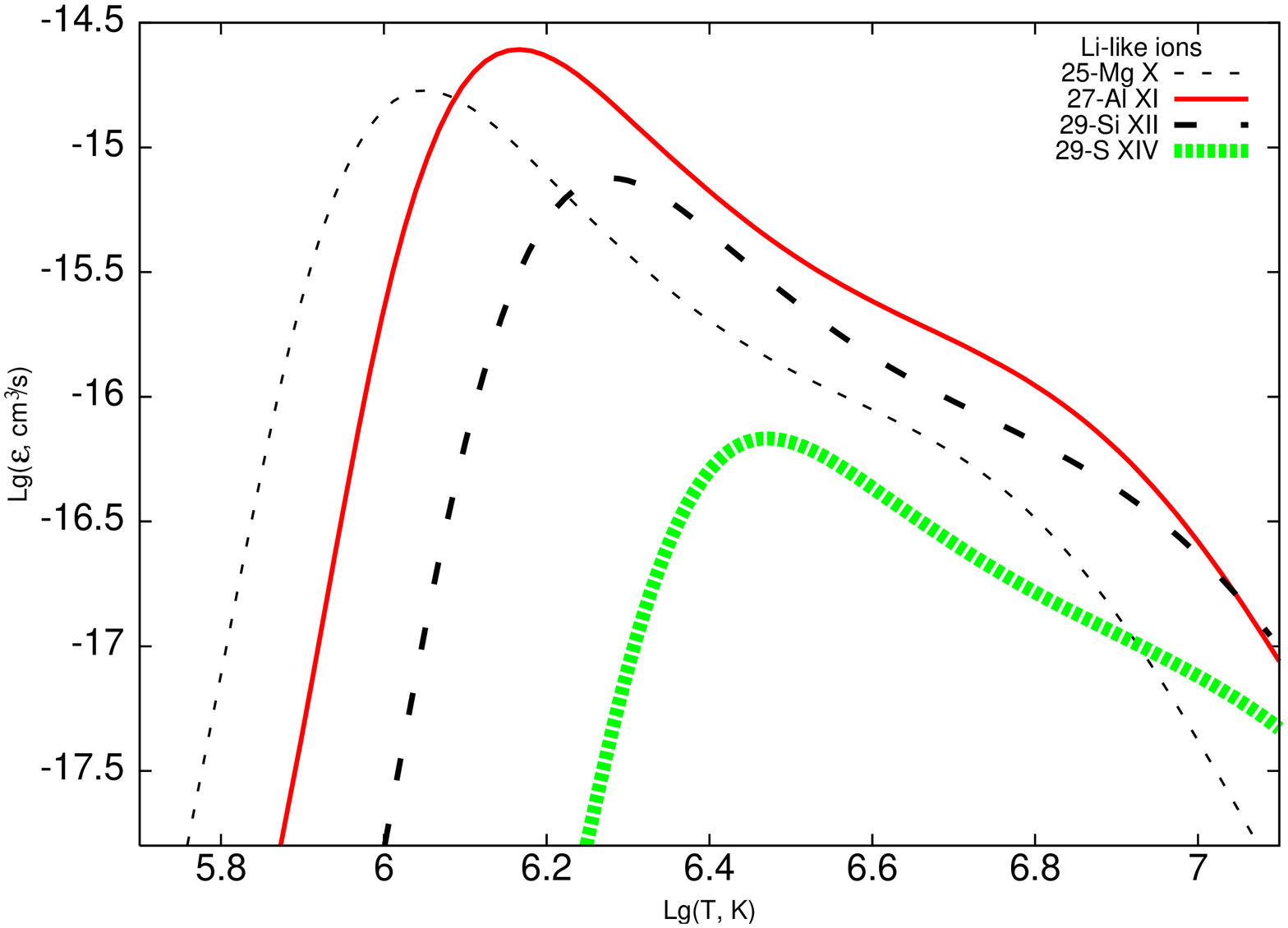}
{{}}              }
  \caption{Emissivity $\varepsilon$, cm$^3$/s, of hyperfine structure 
  transitions of discussed ions.
  Curves are valid for negligibly low electron density and 
  take into account CMB radiation field at redshift $z=0$. 
  Left and right panels present data on H-like and Li-like ions 
  respectively.}
  \label{FigEmiss}
  \end{center}
\end{figure*}

Spectral line emissivity was then computed as
$$
\varepsilon = D(T_R,n_e) X_{ion}(T)\,X_{iso}\,C_{lu}(T),
$$
where 
$D(T_R,n_e)$ is the correction factor defined below,
$X_{ion}(T)$ is ionic abundance (taken from \cite{Mazzotta}), 
$X_{iso}$ is absolute isotopic abundance relative to hydrogen
and $C_{lu}(T)$ is the excitation rate coefficient 
from level $l$ to level $u$, obtained either
from excitation cross-sections or from collision strengths.
Resulting line emissivities as functions of temperature
are shown on Figure~\ref{FigEmiss}.

{{}}

\subsection{Optical depth and emissivity corrections}
\label{SSecCorr}

Computing line emissivity as well as line optical depth,
one has to account for change in 
the hyperfine sublevel population due to interaction with 
cosmic microwave background (CMB) radiation and 
collisional processes.
Both these effects diminish spectral line emissivity 
and line optical depth (see below).

\begin{figure}[htb]
  \begin{center}
{{}}  \centerline{ 
       \includegraphics[width=\linewidth]{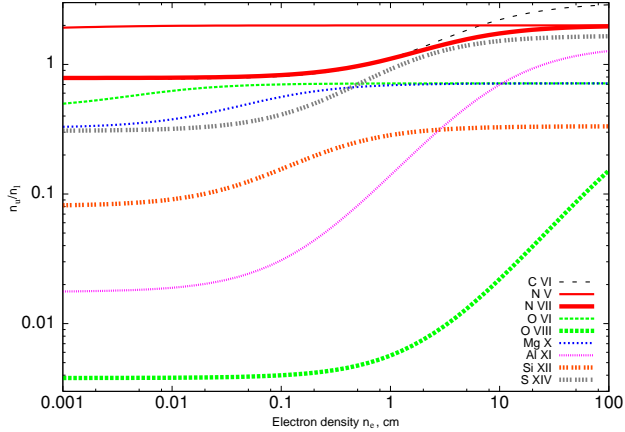}
{{}}             }
  \caption{Upper-to-lower hyperfine sublevel population
  ratio $n_u/n_l$ as a function of electron number density $n_e$
  at $z=0$.
  Behavior in the limiting case of low $n_e$ is determined by CMB
  radiation field (see Eq.~(\protect\ref{EqCMBCorr})).
          }
  \label{FigHFSnunl}
  \end{center}
\end{figure}

If we characterize external radiation fields with effective temperature
$T_R$ at the frequency of transition (in the majority of cases
it will be equal to the CMB temperature\footnote%
{%
In the vicinity of strong radio sources this effective temperature
will be somewhat higher. For the specific sources we are considering, 
the highest effective temperature is in the Cas~A supernova remnant
additionally contributing at wavelength 6.5~mm about 1~K to the 
CMB~\citep{CasAFlux}.
}),
the stationary upper-to-lower hyperfine sublevel population 
ratio is expressed as (e.g., \cite{DCruzSarazin,Liang})
\begin{equation}
  \label{EqNSimple}
  \frac{n_u}{n_l} = \frac{g_u}{g_l}
		\frac{N+\frac{n_e}{n_{cr}}}{1+N+\frac{n_e}{n_{cr}}},
\end{equation}
where $n_e$ denote the electron number density, and
the photon occupation number is
$N=\left[\exp\left(\frac{h\nu}{kT_R}\right)-1\right]^{-1}$. 
Here 
$$
n_{cr} \equiv \frac{A_{ul}}{C_{ul}} = 
              \frac{g_u}{g_l} \cdot \frac{A_{ul}}{C_{lu}}
$$
define critical electron number density. Its values
for HFS transitions are given in Table~\ref{TabHFSCorr}
for isotopes with $n_{cr}<500$~cm$^{-3}$.
HFS level population ratio $n_u/n_l$ as a function of electron number
density is shown in Figure~\ref{FigHFSnunl}.

Then the intensity and optical depth multiplicative correction factor is
\begin{equation}
\label{EqDcorr}
 D(T_R, n_e) = \frac{1}
  {1+\left(1+\frac{g_u}{g_l}\right) \left(N + \frac{n_e}{n_{cr}}\right)}.
\end{equation}

\begin{figure*}[htb]
  \begin{center}
{{}}  \centerline{
{{}}       \includegraphics[width=0.49\linewidth]{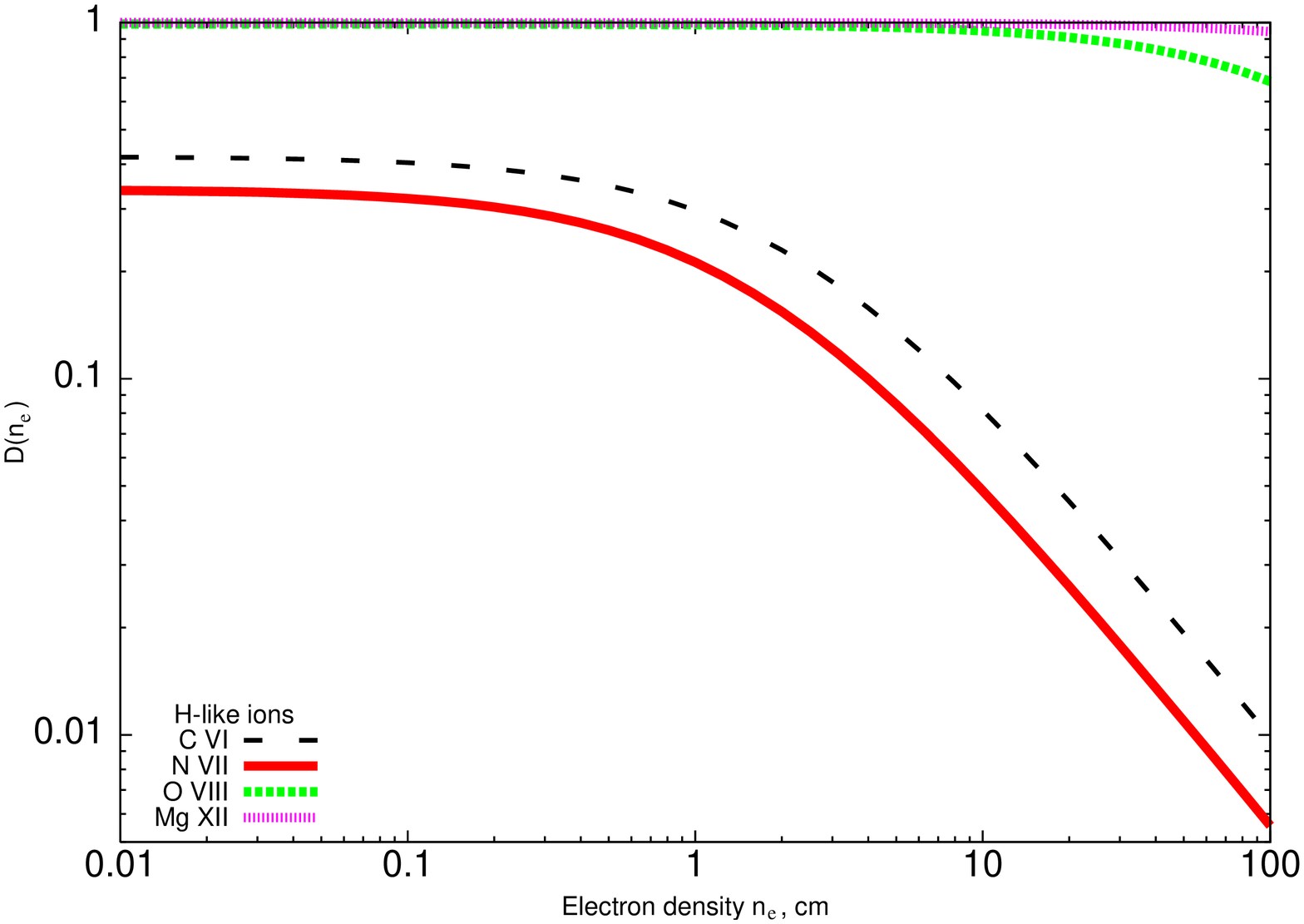}
{{}}       \includegraphics[width=0.49\linewidth]{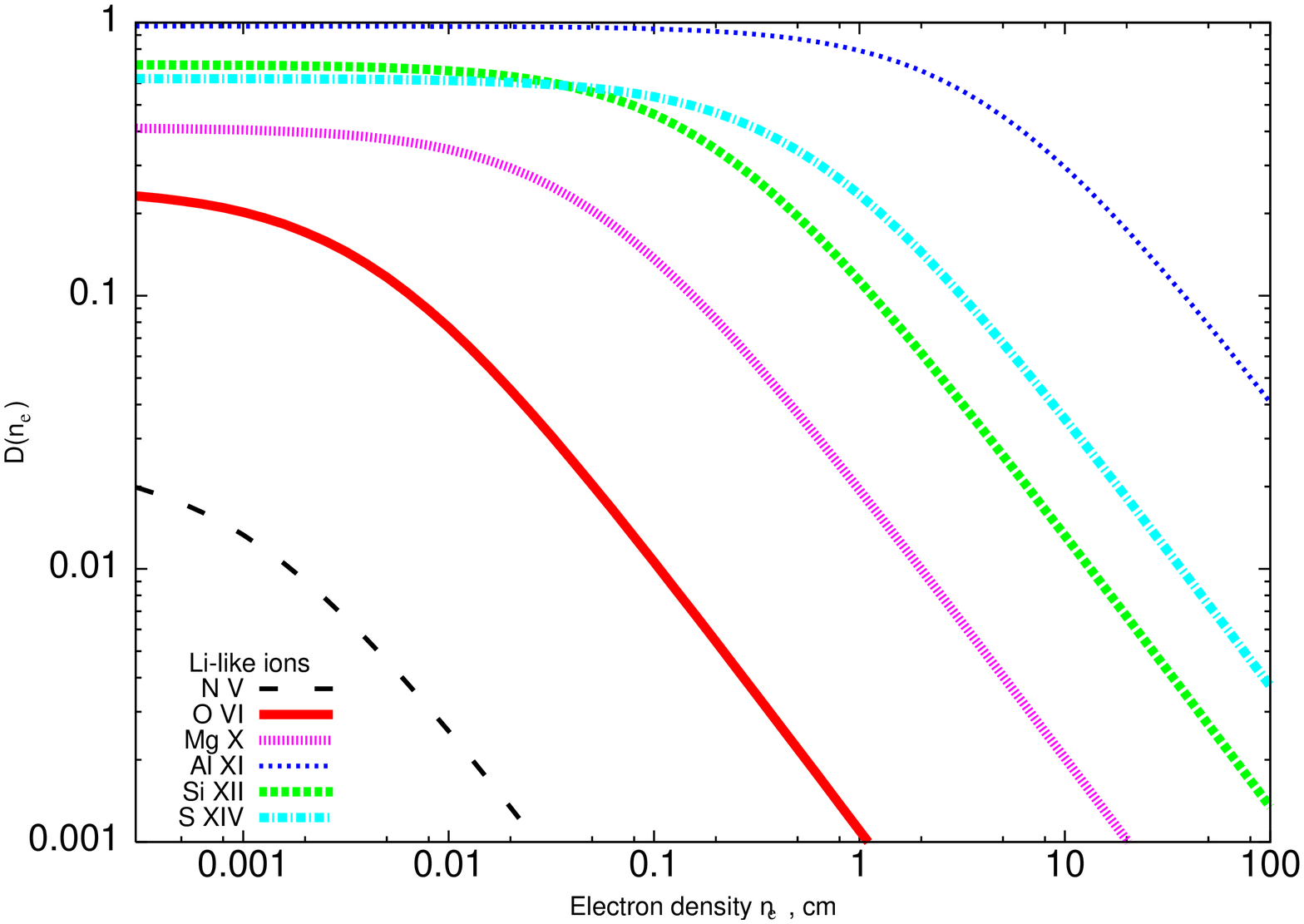}
{{}}              }
  \caption{\label{FigHFSDcorr} Correction coefficient $D(T_{R0},n_e)$
  as a function of electron density $n_e$ for zero redshift, i.e.\ 
  for $T_{R0}=2.725$~K.
  Left and right panels present data on H-like ions and Li-like ions 
  respectively.}
  \end{center}
\end{figure*}

\begin{figure}[htb]
  \begin{center}
  \centerline{ 
    \rotatebox{270}{
       \includegraphics[height=\linewidth]{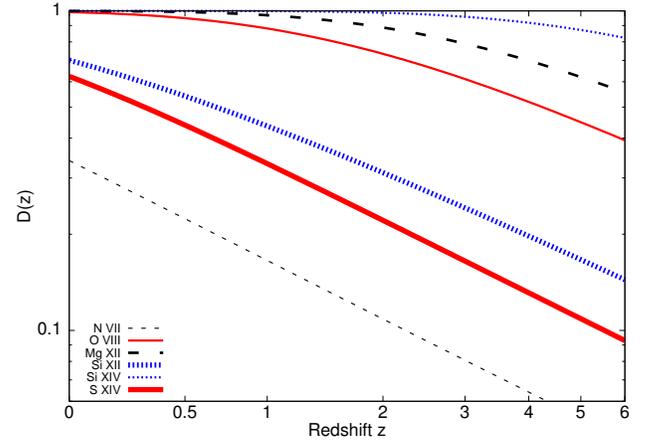}
                   }
             }
  \caption{\label{FigHFSDcorr_z} Correction coefficient $D(T_R,n_e)$
as a function of redshift $z$ for some HFS transitions.
The horizontal scale is linear in $\lg(1+z)$.}
  \end{center}
\end{figure}

In case of $n_e\ll n_{cr}$ this expression reduces to
\begin{equation}
\label{EqCMBCorr}
  D(T_R,0) = \frac{1-\exp(-h\nu/kT_R)}{1+\frac{g_u}{g_l}\exp(-h\nu/kT_R)}.
\end{equation}
Correction coefficients for zero redshift are given in 
Table~\ref{TabHFSCorr} and Figure~\ref{FigHFSDcorr}, 
using $T_{R0}=2.725$~K. We also present correction coefficient
as a function of redshift $z$ for some of relevant transitions on
Figure~\ref{FigHFSDcorr_z}. 

In the opposite case of $n_e\gg n_{cr}$
\begin{equation}
\label{EqDInf}
  D(T_R, n_e\to\infty) = \frac{n_{cr}}{n_e} \cdot
         \frac{1}{1+g_u/g_l},
\end{equation}
hence the absorption line optical depth depends
only on the path length and not on the actual density value.
This means that there is a lower limit on absorbing medium 
path length $l_{min}$, that may give rise to a given optical depth $\tau$. 
Its approximate values for $\tau=0.1\%$ are given in Table~\ref{TabHFSCorr}.
To determine it, we assume
solar elemental abundances and maximum ionization fractions 
(respectively, 0.3 for Li-like and 0.5 for H-like ions).
From this Table it can be seen immediately that only larger objects, such as
WHIM filaments or large amounts of hot ISM, are observable in absorption.
We note that intensities of $^{14}$N V and $^{17}$O VI
hyperfine lines are strongly diminished due to small critical 
densities of corresponding transitions.

The same effect will make the emission line intensity 
in high-density environments to be proportional not
to emission measure $\int n_e^2 dl$, but to $\int n_e n_{cr} dl$, 
thus relatively diminishing contribution from the denser regions. 
This allows one in principle to separate highly-charged ion line emission
of the heliospheric and geocoronal charge-exchange reactions
from the Galactic halo or Local Bubble emission, that is now 
virtually impossible for soft X-ray observatories
(see also below).

Emissivity of HFS lines of Li-like ions, from the other side, can be enhanced
by the resonance excitation by the fine structure lines of the
same ions~\citep{Field21cm,SCh84} situated in the far ultraviolet and
having large resonance scattering cross-sections.

{{}}\begin{table}[thb]
\caption{\label{TabHFSCorr}Radiative correction coefficients 
(for redshift $z=0$) 
and critical density values (at $T=10^6$~K) for 
relevant HFS transitions. The last column contains minimum path
length necessary for achieving optical depth of 0.1\%.}
\begin{center}
\begin{tabular}{|l|l l c|}
\hline
Isotope, ion & $D(T_{R0},0)$  & $n_{cr}$, cm$^{-3}$ & $l_{min}$, kpc\\
\hline
$^{13}$C VI  & 0.421        &  4.1    & 0.3 \\

$^{14}$N V   & 0.0252       & $8.5\cdot10^{-5}$ & 900  \\

$^{14}$N VII & 0.340        &  1.7    & 0.03 \\

$^{17}$O VI  & 0.247        &  0.0017 & 9000 \\

$^{17}$O VIII& 0.991        &  380    & 0.03  \\

$^{25}$Mg X  & 0.413        &  0.034  & 50   \\

$^{27}$Al XI & 0.970        &  10.2   & 0.10 \\

$^{29}$Si XII& 0.701        &  0.18   & 25  \\

$^{33}$S  XIV& 0.624        &  1.77   & 18  \\
\hline
\end{tabular}
\end{center}
\end{table}

{{}}

{{}}

{{}}

\section{Emission lines from hot ISM and supernova remnants}
\label{SecEmission}

\subsection{Overview of the brightest objects}
\label{SSecObj}

To be bright in hyperfine structure emission lines, the plasma should
have high emission measure and appropriate temperature. Besides,
among objects of equal emission measure the \textit{least dense} 
will be the brightest due to diminishing of the $D(T_R,n_e)$ with density,
as described in the previous section.

Therefore the main types of objects
with expected bright emission lines
are young and middle-aged supernova
remnants (SNR) and the hot interstellar medium (ISM) including galactic
halos. As the same H-like ions that we are considering have spectral lines
in the soft X-ray band, the objects with bright X-ray lines should also be 
bright in HFS lines.

According to these selection criteria, we have chosen several objects
with parameters given in the Table~\ref{TabHFSObj}. They include 
Vela XYZ~\citep{LuVela},
Cygnus Loop~\citep{CLoopROSAT} and Cassiopeia~A~\citep{CasAChandra} 
Galactic supernova remnants, bright
supernova remnant N157B~\citep{N157B} in Large Magellanic Cloud and
hot interstellar gas in the Local Hot Bubble, cool Galactic halo and
North Polar Spur~\citep{MWHalo}. All of the bright spots considered
in these objects are diffuse 
(i.e.,\ larger than or comparable to the radio telescope angular resolution),
therefore we use the notion of differential brightness temperature 
$T_b$ in our analysis.

{{}}\begin{table*}[tbh]
\caption{Objects expected to be bright in HFS lines and discussed
further include several supernova remnants, such as 
Vela XYZ, N157B, Cygnus Loop (CL) and Cassiopeia~A and 
some galactic objects such as as Local Hot Bubble (LHB), 
cool Galactic halo (CGH) and North Polar Spur (NPS). Rough estimates for
hot interstellar medium (HISM) in the plane of the Milky Way is also given.
Table includes distance to the object $d$, 
bright fragment angular size $\theta$, 
its average electron density $n_e$ and temperature $T_e$, 
linear extent $l$, emission measure $n_e^2l$ and equatorial
J2000 coordinates.}
\label{TabHFSObj}
\begin{center}
\begin{footnotesize}
\begin{tabular}{|l|r r r r r r r r|}
\hline
            & LHB    & CGH    & NPS   & Vela  & N157B & CL   &Cas~A & HISM\\
\hline
$d$, kpc    &  ---   & ~0.4   & ~0.1  & 0.25  & 50    & 0.44 & 3.4  & --- \\

$\theta$    & 30'    & 30'    & 30'   & $>5'$ & 1'    & 5'   & 5''  & --- \\

$n_e$, cm$^{-3}$ 
            & 0.010  & 0.010  & 0.031 & 0.5   & 3.8   & 5.5  & 160  & 0.01 \\

$kT_e$, keV & 0.10   & 0.10   & 0.26  & 0.12  & 0.23  & 0.12 & 0.77 & 0.10 \\

$l$, pc     & 60     & 1000   & 160   & 8     & 14    & 1.2  & 0.08 & 3000 \\

RA, J2000   
 &16$^{\rm h}$42$^{\rm m}$ &16$^{\rm h}$26$^{\rm m}$& 16$^{\rm h}$26$^{\rm m}$
 & 08$^{\rm h}$57$^{\rm m}$  & 05$^{\rm h}$37$^{\rm m}$54$^{\rm s}$
 & 20$^{\rm h}$50$^{\rm m}$ & 23$^{\rm h}$23$^{\rm m}$35$^{\rm s}$ & --- \\

Dec, J2000 
 &  02$^\circ$19'  &  03$^\circ$11'  & 03$^\circ$11'  
 & -45$^\circ$00'  & -69$^\circ$09' 50''
 & 32$^\circ$11'   &  58$^\circ$50'05'' & --- \\

\hline 
$n_e^2l$, cm$^{-6}$ pc
            & 0.005  & 0.088  & 0.13  & 2.2   & 200 & 35   & 1000 & 0.3 \\
\hline
\end{tabular}
\end{footnotesize}
\end{center}
\end{table*}

Young supernova remnant Cassiopeia~A (Cas~A) stands separate in this list
as its brightest regions contain material strongly enriched in the 
supernova explosion. In the considered regions 
R1, R3 and R4 from \cite{CasAChandra} the oxygen ions are 
dominant (constituting more than $80-90$\% by mass), 
hence the abundances of 
oxygen, magnesium, silicon and sulphur isotopes of interest 
may be about three orders of magnitude higher than in other cases, 
if making rather natural assumption of Earth isotopic mole fractions.
This is the main reason of high-intensity
signal coming from Cas~A (see Table~\ref{TabHFSEmission}).
Detection of these isotopes will give important information
about nuclear processes leading to formation of isotopes of
$\alpha$-elements before and during a supernova explosion.

We use a simple model to estimate emission and absorption
in the hot interstellar medium (HISM) of the Milky Way. For this
we assume the same temperature and density as in the Local Hot Bubble
(see Table~\ref{TabHFSObj}), but take the path length of 3 kpc
(from the total path through the Galaxy of 15~kpc and hot 
ISM filling fraction of 20\%).

\subsection{Emission line intensity estimates}

From the beginning let us discuss the $^{14}$N~VII emission line arising
in distant ($z>0$) sources.
First let us estimate a possibility of its detection
from halo of a spiral galaxy. 
Assuming Milky Way parameters (see above), 
galactic halo volume emission measure is 
$n_e^2V\approx1\cdot10^{63}$~cm$^{-3}$. HFS line emissivity for temperature 
$T\approx1.5\cdot 10^6$~K is $3\cdot10^{-15}$~cm$^3$/s 
(see Figure~\ref{FigEmiss}). 
Assuming solar nitrogen abundance,
the luminosity in the $^{14}$N~VII spectral line is about 
$1\cdot10^{33}$~erg/s.

Another sources of the $^{14}$N~VII HFS emission are supernova remnants 
(this has been suggested already by \cite{SCh84}) at age
between one and ten thousand years. As a numeric example, let us
take bright SNR Puppis~A~\citep{PuppisAChandra}. Its studies indicate 
hot gas emission measure of $n_e^2V\approx2\cdot10^{59}$~cm$^{-3}$. 
Accounting for some diminishing of the correction factor, the
luminosity in the spectral line is about $1\cdot10^{29}$~erg/s.

Observing the polar cap of nearby galaxy M82 having high 
star formation rate (3.6 solar masses per year) and 
apparently being in the process of merger,
\textit{Suzaku} and \textit{XMM-Newton} orbital telescopes have
discovered the gas in the temperature range $(2-6)\cdot10^6$~K
\citep{M82Cap} optimal for the  $^{14}$N~VII line detection.
It is argued that this gas is moving from the galaxy with
hyperbolic velocity.

In a starburst galaxy the star formation rate may be as high as
500 solar masses per year or even higher. 
This corresponds to supernova rate of
about ten per year, therefore the total number of supernova of age below
$10^4$~years should be several tens of thousands. 
Note that a lot of gas reside in a starburst galaxy, hence
supernovae will explode in dense medium forming a lot of bright 
high density supernova remnants.
In such galaxy 
the total luminosity in line may be as high as $10^{34}-10^{35}$~erg/s.

At the redshift of $z=0.15$ luminosity in line of $1\cdot10^{34}$~erg/s 
corresponds to signal of only about $3\;\mu$Jy in the line of about
40~km/s width, showing that the emission 
signal from this class of objects is difficult to detect using 
existing instruments. Though, the next generation instruments such as
\textit{SKA} should be able to observe it from high-redshift 
($z>1.1$) star-forming galaxies.

As the zenith atmospheric transmission at 53~GHz at the height of 
Chajnantor plane reaches up to 50\%, telescopes such as \textit{APEX}
and, later, \textit{ALMA} will be
able to conduct the first observations of the $^{14}$N~VII line
from supernova remnants such as N157B which is especially bright
in this spectral line.

\bigskip
In case of observations from Chajnantor plane or in case of
of emission lines in atmospheric transparency bands, 
it is preferable to observe first the brightest sources of our Galaxy.
In this case the received flux will accordingly rise.
Estimates of brightness temperature $T_b$ in hyperfine structure lines 
from mentioned objects of the Galaxy and its
neighborhood are given in Table~\ref{TabHFSEmission}.
Solar elemental abundances are assumed everywhere, except for
O, Mg, Si and S ions in Cas~A supernova remnant, where they are
known from \cite{CasAChandra}.
Differential brightness temperature of the same order of magnitude
is expected as long as the observed object (for example, a galaxy) 
is larger than the angular resolution of the radio telescope.

We estimate $T_b$ in 
the $^{25}$Mg~X HFS line arising in the Milky Way halo, Vela, N157B
and Cygnus Loop supernova remnants to be about 5--15~$\mu$K. As 
another example, in younger and hotter oxygen-rich Cas~A supernova 
remnant the brightness temperature in mm-band 
$^{27}$Al~XI, $^{29}$Si~XII and $^{33}$S~XIV
hyperfine transition lines is estimated to constitute about 
40--80~$\mu$K.
Accounting for resonance excitation due to significant optical depth
in the Si~XII and S~XIV ionic fine structure transitions in the
bright Cas~A fragments, HFS lines of isotopes of these ions will
be enhanced by factors of 1.5 and 2, respectively (in the 
Table~\ref{TabHFSEmission} we give values accounting for this
increase). For other described sources there is no strong
enhancement of HFS line intensity (for example, in N157B 
the resonance excitation contributes additionally about 20\%,
assuming solar elemental abundances).

Sub-millimeter lines of $^{25}$Mg~XII and $^{29}$Si~XIV arising in
Cas~A are of the same order of brightness reaching about 50~$\mu$K.
Unfortunately, there is significant atmospheric obscuration on 
frequencies of these lines at zero redshift, that may give
rise to difficulties in the ground-based detection. Simple
estimates of \textit{ALMA}
sensitivity, taking its system temperature from sensitivity calculator%
\footnote{http://www.eso.org/projects/alma/science/bin/sensitivity.html}
to be 1600~K give 3-$\sigma$ detection limit on the order of 0.4~mK
at 658~$\mu$m (wavelength of $^{25}$Mg~XII line) that is about one
order of magnitude above the predicted emission line intensities.

Estimates of emission line intensity from WHIM filaments, similar 
to discussed above, give differential brightness temperature
significantly below 1~$\mu$K due to their low emission measure.

Most of the bright spots we are discussing have angular sizes
of the order of arcminute that is similar to the angular resolution
of PSPC detector aboard \textit{ROSAT} and of a 10-meter-class
radio telescope.
Computing total luminosity of a bright spot it is sometimes comfortable to
use its flux $S$ that is connected with the brightness temperature
from Rayleigh-Jeans formula as 
$$
S = 0.23
     \cdot \left(\frac{T_b}{\mbox{1 $\mu$K}}\right)
     \cdot \left(\frac{\lambda}{\mbox{1 mm}}\right)^{-2}
     \cdot \left(\frac{\theta}{\mbox{1'}}\right)^2 \mbox{ mJy},
$$
where $\theta$ denote the brightest spot angular size.

{{}}\begin{table*}[tbh]
\caption{\label{TabHFSEmission}Brightness temperature $T_b$, $\mu$K,
in hyperfine structure emission lines from the bright fragments
of objects from the Table~\protect\ref{TabHFSObj}.
Values less than 0.01~$\mu$K are not shown.
Unless noted otherwise, solar elemental abundances are assumed.
}
\begin{center}
\begin{tabular}{|l l|r r r r r r r r|}
\hline
Isotope  & $\lambda$, mm 
 &  LHB   & CGH    & NPS   & Vela & N157B$^*$& CL  & Cas~A    & HISM\\
\hline
$^{13}$C VI & 3.8740(8) 
 & 0.01   & 0.15   & ---   & 1.7  & 4   & \fbox{13} & --- & 0.4 \\

$^{14}$N VII & 5.6519(11) 
 & 0.2    & 3.5    & 1.0   & \fbox{70} & \fbox{700} & 
                                     \fbox{400} &  7 & \fbox{20}\\

$^{17}$O VIII& 1.0085(2)
 & ---    & ---    & ---   & 0.01 & 0.8 &0.08&  8$^*$       & ---\\

$^{25}$Mg X  & 6.680(4)
 & 0.3    & 5.1    & 0.4   & 8.0  & \fbox{15}&\fbox{14} & 1.0$^*$ & \fbox{15}\\

$^{25}$Mg XII& 0.65809(13)
 & ---   & ---    & ---    & ---  & ---& --- &\fbox{40$^*$} & --- \\

$^{27}$Al XI & 1.2060(7)
 & 0.01  & 0.1    & 0.06   & 4.2  & \fbox{50} & \fbox{40}
                                                    &1   & 0.8 \\

$^{29}$Si XII& 3.725(2) 
 & ---    & 0.04   & 0.30  & 1.2  & \fbox{30}&3.0 &\fbox{120$^*$} & 0.7 \\

$^{29}$Si XIV& 0.38165(7)
 & ---   & ---    & ---    & ---  & --- & --- &\fbox{40$^*$} & --- \\

$^{33}$S  XIV& 3.123(2)
 & ---    & ---    & 0.15  & ---  & 10  & --- &\fbox{150$^*$} & --- \\

\hline
\multicolumn{10}{l}{$^*$ elemental abundances are known from observations}
\end{tabular}
\end{center}
\end{table*}


\subsection{Resonant scattering in the surroundings of quasar}

Detection of emission lines arising in the process of the
resonant scattering of the quasar radio emission in the
hot and warm gas surrounding the quasar 
\citep{QuasarSpectrum,Kallman1982}
might be especially interesting.
Discovery of a line formed by the resonantly scattered radiation
may shed light upon the quasar radiation beam width, 
as for the narrow beam much less gas is irradiated
\citep{Cramphorn2004}.

As in the quasar neighborhood the photon density 
might highly exceed one of the CMB radiation, 
the correction coefficient will decrease there (see Eq.~\ref{EqDcorr}).
Despite this, arising line brightness is of the same order of 
magnitude as of the lines described above.

We show it on example of 3C~273 ($z=0.158$) quasar, having
flux about 30~Jy on the wavelength of $^{14}$N~VII ion line.
The quasar radiation photon occupation number at the line frequency
exceeds the one due to the CMB inside central 70~kpc.
Even so, despite the very low average value of $D\approx3\cdot10^{-3}$,
the additional flux due to resonant scattering in the line center
will constitute about 0.5~mJy
(assuming gas cloud parameters similar to ones observed around 
elliptical galaxies).
A large number of young supernova remnants is expected to exist
in the central region of galaxy surrounding a quasar.
Gas of these SNRs will resonantly scatter quasar radiation,
additionally increasing the signal.

\subsection{Disentangling Local Bubble and heliospheric emission}

One of serious problems in studying the Local Bubble --- the hot gas 
cavity around the Solar System --- is difficulty of
separation of its soft X-ray line radiation from foreground contribution
arising in the solar wind and the Earth corona 
(respectively, heliospheric and geocoronal emission). 
The latter arise in charge-exchange 
collisions~\citep{CoxCXE} that are so effective that intensity of e.g. O~VIII
line may significantly exceed one of the Local Bubble~\citep{MBM12Chandra}.

There is an intrinsic difficulty in disentangling these two spectral line
contributions in soft X-rays, as both these sources are diffuse.
Though, using simultaneous observations of soft X-ray and HFS lines it
becomes possible 
to separate them, as the HFS 
line emissivity is dependent on density (see Figure~\ref{FigHFSDcorr})
and for some ions the correction coefficient is significantly different in 
dilute Local Bubble gas ($n_e\approx0.01$~cm$^{-3}$), 
denser solar wind ($n_H\approx0.1$~cm$^{-3}$)
and geocorona ($n_H\approx10$~cm$^{-3}$).
In the last two cases we give the neutral hydrogen number density as 
it is the primary source of electrons participating in the 
charge-exchange collisions.

As a tool for such density diagnostics one may take $^{25}$Mg~X line
having critical density of about 0.03~cm$^{-3}$, that is less than
electron density in the Solar System. Hence its line emissivity will
be suppressed by a factor of roughly $n_e/n_{cr}$, this factor being
dependent on the distances to Sun and to Earth.

For estimates of separate source contributions we use simple models
of neutral hydrogen distribution in heliosphere and geocorona and 
rates of charge-exchange reactions from~\cite{CXEsigma} and \cite{CXEparams}. 
It follows that the main part of heliospheric emission arises
at distances larger than 1~a.u.\ where the solar wind densities fall
down to about $0.1-1$~cm$^{-3}$. Hence $^{25}$Mg~X line emissivity
in heliosphere is suppressed by a factor of several (about three) as
compared to the Local Hot Bubble.
Geocoronal emission mostly arises in regions where electron
density is much more than $1$~cm$^{-3}$. Therefore its $^{25}$Mg~X 
line emission is suppressed much stronger (by factor of about 100)
and is virtually unobservable.

The practical problem of this method, however, is an extremely 
low brightness temperature from each of these sources. 
As indicated in Table~\ref{TabHFSEmission}, expected
$T_b$ from the Local Bubble is below 1~$\mu$K; expected contribution
from the Solar System is of the same order of magnitude.

\section{Absorption lines in WHIM and hot ISM}
\label{SecAbsorption}

\subsection{Warm-hot intergalactic medium}

The prospects of detection of highly-ionized oxygen O~VII and O~VIII
ion absorption X-ray lines from the WHIM are being widely discussed 
now (e.g., \cite{Gnedin,ChenWHIM,CenWhere3}). We estimate below the
feasibility of detection of the $^{14}$N~VII ion HFS absorption line from 
this medium. Only this isotope is considered, as other HFS absorption lines
(e.g., of $^{25}$Mg~X, $^{29}$Si~XII) 
will be at least one order of magnitude weaker
due to smaller isotopic abundances (see Figure~\ref{FigHFSabund}).

Absorption cross-sections of O~VII and O~VIII X-ray transitions around 
20~\AA\ are $\sigma\approx(2-5)\cdot10^{-16}$~cm$^2$,
i.e.\ about three orders of magnitude larger than of HFS transitions.
Moreover, optical depth $\tau$ of HFS transition is additionally 
diminished due to population of upper hyperfine sublevel in the
field of the CMB radiation. 
Resulting rough estimate of $^{14}$N~VII HFS line optical depth 
corresponding to O~VII or O~VIII soft X-ray line $\tau{\rm (O)}\approx1$
is only about $\tau{(\rm ^{14}N VII)}\approx (3-10)\cdot10^{-5}$.

Note that weak emission lines of comparable magnitude have 
already been detected on \textit{GBT}. As an example, HCN molecular 
line with line width of $\Delta v=140$~km/s was observed by \cite{HCNonGBT} 
in emission from galaxy at $z=2.28$ at frequency
27.0~GHz with 1-$\sigma$ flux uncertainty of 0.1~mJy.
This flux corresponds to optical depth of only $1\cdot10^{-5}$, 
if observed from the source with flux of 10~Jy.

To assess frequency of occurrence of $^{14}$N~VII absorption lines in WHIM,
we use the distribution function of O~VIII X-ray absorption 
line equivalent width from~\cite{CenWhere3} and correspondence between
the equivalent width and ionic column density from \cite{ChenWHIM}
cosmological simulations.
As a first approximation, we also assume that in WHIM conditions
the ionization equilibrium curve of O~VIII atoms 
is the same as of N~VII atoms\footnote{%
In collisional ionization equilibrium conditions, N~VII is abundant 
at slightly 
lower temperatures than O~VIII. 
But in the low-density environments ionic abundances change due to 
photo-ionization and, as a result, N~VII abundance widens 
and becomes rather similar to one predicted for O~VIII ion~\citep{ChenWHIM}. 
%
}
and their relative abundance N/O is solar. 
Then on average in one sight line in the redshift intervals 
$z=0.15-0.30$ and $z=0.3-0.6$ there is expected one $^{14}$N~VII 
HFS line with $\tau\ge 2\cdot10^{-5}$ and $\tau\ge 3\cdot10^{-5}$,
respectively.
In case of reported N~VII soft X-ray line 
detection by~\cite{Nicastro} with ionic column density of 
$1.5\cdot10^{15}$~cm$^{-2}$
the optical depth in absorption HFS line would constitute about
$\tau=3\cdot10^{-4}$.

\subsection{Hot interstellar medium}

Another case of interest concerns absorption by the hot ISM of
a spiral galaxy with strong radio source in it. 
For example, every several years exceptionally luminous radio sources
(so-called microquasars) burst in the Milky Way.
They retain their outstanding brightness for the time period 
of several weeks and are thus good candidates to 
study absorption lines in the hot ISM.

As a specific example of a luminous outburst, there has been
observed a brightening of Cygnus~X-3 with flux as high as 20~Jy at
10.5~GHz~\citep{1972CygX-3} and spectral index estimate of $\alpha\ge-0.5$. 
Cygnus~X-3 lies in the galactic plane at a distance of about 9~kpc.
Taking the Milky Way parameters as discussed above,
the column density of hydrogen ions in hot ISM in the direction 
of Cygnus~X-3 is about $6\cdot10^{19}$~cm$^{-2}$. Resulting HFS line 
optical depths and respective absorption line (negative) fluxes 
are given in Table~\ref{TabCygX3}.

{{}}\begin{table}[thb]
\caption{\label{TabCygX3} HFS absorption line optical depth $\tau$
estimate from Cygnus~X-3. The last column contains corresponding 
absorbed flux assuming source brightness of 20~Jy at 10~GHz and spectral
index of $\alpha=-0.5$.
Only lines with resulting optical depth of more than $3\cdot10^{-6}$
are shown.
}
\begin{center}
\begin{tabular}{|l r | r c|}
\hline
Isotope, ion & $\lambda$, mm & \multicolumn{1}{c}{$\tau$} 
                               & $-S_{abs}$, mJy  \\
\hline
$^{13}$C VI  & 3.8740(8)  & $3\cdot10^{-5}$ & 0.17 \\

$^{14}$N VII & 5.6519(11) & $3\cdot10^{-4}$ & 2.0  \\

$^{25}$Mg X  & 6.680(4)   & $1\cdot10^{-5}$ & 0.10 \\

$^{27}$Al XI & 1.1060(7)  & $1\cdot10^{-5}$ & 0.04 \\

\hline
\end{tabular}
\end{center}
\end{table}


\subsection{Gamma-ray burst afterglows}

Significant part of bright gamma-ray afterglows originates from the
star-forming regions or star-forming galaxies. The lifetime of afterglow
in radio band is longer than days or even weeks. This opens a unique
possibility to measure absorption lines in their spectra connected with 
hot ISM of the parent galaxy, 
hot intracluster medium of the corresponding cluster of galaxies (if any)
and the WHIM along the whole path of radio waves to the source.

The optical depth of the absorption features should not diminish
significantly with redshift. Though, the flux from gamma-ray burst 
radio afterglows (several mJy) is weaker than of brightest high-$z$
quasars by about three orders of magnitude that makes the detection
of HFS absorption lines a task for the next generation of interferometers
such as \textit{SKA}.

\subsection{Estimates of HFS line detectability}

Now let us estimate a modern telescope abilities on example of the 
\textit{Green Bank Telescope} (\textit{GBT})
observing absorption line from 3C~279 -- one of the the strongest radio 
sources at 40~GHz with flux of about 15~Jy -- with integration time 
10~hours and frequency resolution of 1~MHz.

As an example, we assume line redshift of $z=0.3$ so that 
the observed $^{14}$N~VII
line is shifted from the atmospheric absorption region.
The \textit{GBT} telescope system temperature near 40~GHz is about 75~K, 
therefore during the
integration time the 1-$\sigma$ flux RMS is on the order of 0.1~mJy. Hence,
neglecting systematic effects, 
the 3-$\sigma$ detection limit of the optical depth will be of the
order of $0.3\,{\rm mJy}/15\,{\rm Jy}\approx2\cdot10^{-5}$ which is 
of the same order as the N~VII HFS line optical depth estimates in 
WHIM or galactic halo.

As another example, we estimate detection feasibility of
the same line in the spectrum of high-$z$ quasar 2134+004
($z = 1.934$). Its flux density on 18~GHz (redshifted frequency
of $^{14}$N~VII) line is about 5~Jy that allows to detect
with similar observation parameters (10 hours integration time, velocity
resolution 10~km/s) 
absorption lines of $\tau\approx6\cdot10^{-5}$ on 3-$\sigma$ level.

From this we conclude that the 5.65~mm $^{14}$N~VII
hyperfine structure line 
in redshifted WHIM filaments might be detectable in absorption 
using existing ground-based instruments.
Searching for it, it is natural to start from the
extragalactic sources brightest in mm band
and having redshift $z>0.15$, such as 
3C~273, 3C~279, 3C~345, 3C~454.3 and 2145+067.


\bigskip
Two more notes would be helpful here. First, from observations
of one absorption spectral line it is difficult 
to both determine redshift and identify the line. 
Hence observations of known objects or
combined with other wavelength data would be desired. 
Second, HFS absorption line 
with the same optical depth is easier to detect at higher 
redshifts, as observed line wavelength moves to lower frequencies 
where the illuminating background source flux is stronger and 
telescope sensitivity becomes higher. Opposing it is the factor 
$D(T_R,n_e)$ that is decreasing 
optical depth in similar gas clouds 
with redshift $z$ for $^{14}$N~VII line
approximately as $(z+1)^{-1}$ (see Figure~\ref{FigHFSDcorr_z}).

\section{Conclusions}
\label{SecResults}

We discuss the feasibility of the
emission and absorption hyperfine structure line observations 
from astrophysical objects in temperature range of $10^5-10^7$~K. 
We find that thanks to $^{14}$N high isotopic abundance in 
interstellar and intergalactic gas the line of $^{14}$N~VII is the
most prospective to be observed in absorption
in spectra of brightest mm-band extragalactic radio sources with $z>0.15$.
Typical optical depth predicted for
WHIM filaments is about $10^{-4}$ that is within the reach 
of the existing instruments. Other HFS lines in absorption will only be
detectable with the next generation radio telescopes.

We estimate the $^{14}$N~VII brightness temperature in emission from 
several Galactic supernova remnants to constitute up to 700~$\mu$K.
Brightness temperature in other lines of $^{25}$Mg, $^{27}$Al, $^{29}$Si
and $^{33}$S reaches up to 100~$\mu$K in oxygen-rich SNR Cas~A,
the level on the limit of sensitivity of the existing instruments
and accessible to future observatories.

Observations of hyperfine structure lines might
provide additional information on isotopic composition of supernova remnants,
help to discover obscured Galactic SNRs, 
allow studying star-forming galaxies and superwinds,
discriminating between heliospheric and Local Bubble
contribution to diffuse soft X-ray background. 
They might also allow to measure hyperfine splitting experimentally
with high precision, thus letting 
to choose between theoretical highly-charged heavy ion 
hyperfine splitting computation models and 
to precise isotopic nuclear magnetic moments.

\section*{ACKNOWLEDGEMENTS}

Authors are grateful to 
I.L.~Beigman, L.A.~Vainshtein and E.M.~Churazov for their
useful remarks 
and to 
B.~Aschenbach, T.~Dotani, R.~Mushotsky and R.~Smith 
for information about the brightest objects of the soft X-ray sky. 


\bibliographystyle{aa}
\bibliography{Bibl_Doc}

\end{document}